\documentclass[aps,prb,longbibliography,floatfix,superscriptaddress,showpacs,amsmath,amssymb,reprint,nofootinbib]{revtex4-2}

\usepackage{graphicx}
\usepackage[english]{babel}
\usepackage{bm}
\usepackage{bbm}
\usepackage{amsmath,amsfonts}
\usepackage{amsbsy}
\usepackage[colorlinks=true,linkcolor=blue,urlcolor=blue,citecolor=blue,breaklinks=true]{hyperref}
\usepackage[utf8]{inputenc}
\usepackage{graphics}
\usepackage{subcaption}
\usepackage{xcolor}
\usepackage{bbm}
\usepackage{graphics}
\usepackage{graphicx}
\usepackage{epsfig}
\usepackage{amsmath}
\usepackage{amssymb}
\usepackage{amsfonts}
\usepackage{bm}
\usepackage{color}
\usepackage{xcolor}
\usepackage{bbm}
\usepackage{hyperref}
\usepackage[normalem]{ulem}
\usepackage{comment}

\begin{document}\title{ 
Low-temperature dissipative conductivity of superconductors with 
paramagnetic impurities
}

\author{Shiang-Bin Chiu}
\affiliation{Department of Physics, University of Washington, Seattle, WA 98195, USA}

\author{Anton Andreev}
\affiliation{Department of Physics, University of Washington, Seattle, WA 98195, USA}

\author{Alexander Burin}
\affiliation{Department of Chemistry, Tulane University, New Orleans, LA 70118, USA}

\author{Boris Z. Spivak}
\affiliation{Department of Physics, University of Washington, Seattle, WA 98195, USA}
\date{\today}

\begin{abstract}
In \emph{s}-wave superconductors with a small concentration of  magnetic impurities, the only electronic excitations that remain available  at low temperatures are the excitations of the system of localized spins. 
We discuss a new mechanism of interaction between electromagnetic waves and the localized spins in disordered superconductors. A supercurrent induces randomly distributed spin density of the itinerant electrons, which couples to the impurity spins by exchange interaction. Acceleration of the Cooper pair condensate by the  external \emph{ac} electric field of frequency  $\omega$ creates a strong, time-dependent exchange field acting on the localized spins, which is inversely proportional to  $\omega$. 
As a result, the low-frequency dissipative part of the conductivity saturates to a nonzero value.  We use the fluctuation-dissipation theorem to evaluate the spectrum of  equilibrium current fluctuations associated with the fluctuation in the spin subsystem.
We also predict that in the presence of a \emph{dc} magnetic field parallel to the superconducting film, the system of spins exhibits a large positive magnetoconductance. 
\end{abstract}

\maketitle

In \emph{s}-wave superconductors at temperatures much lower than the critical temperature, $T\ll T_{c}$, and at sufficiently small concentrations of magnetic impurities, $n_{s}$, the only thermally-active electronic excitations are associated with the system of impurity spins.
In this case, the low frequency dissipative conductivity of the system will be dominated by the interaction of the electromagnetic (EM) field  with the  impurity spins. 

We consider magnetic impurities with half-integer spins, and assume that the crystalline anisotropy lowers the degeneracy of the low-energy manifold of each impurity to the Kramers doublet, $S=1/2$~\footnote{This is a generic situation, see for example $\S$ 99 of Ref.~\cite{landau2013quantum}.}.  
The linear coupling of the impurity spins $\hat{\bm{S}}_a$ is allowed only to those perturbations that break time-reversal symmetry (TRS). This coupling is described by the Hamiltonian of the form
\begin{subequations}
\label{eq:Eq_1}
    \begin{align}
\label{eq:spin_hamiltonian}
        \hat{H}_{\mathrm{S}} (t) = & \,   - \sum_a \hat{\bm{S}}_{a}\cdot {\bf H}_{\mathrm{eff},a} ( t) , \\
        \label{eq:Heff}
{\bf H}_{\mathrm{eff},a}( t) = & \,  \left[  \mathcal{J} \hat{\bm{s}} ({\bf r}_a,t)  + 
\mu_{\mathrm{B}} g_{a}{\bf B} ({\bf r}_a,t)  \right].
    \end{align}
\end{subequations}
Here $\mathcal{J}$ is the exchange constant, $\mu_{\mathrm{B}}$ is the Bohr magneton, and  $g_a$ are $g$-factors of magnetic moment of the impurity which in the disordered system can be spatially nonuniform. The first term in the square bracket describes the exchange coupling with the spin density of itinerant electron $\hat{\bm{s}} (\bm{r}_a) $ at the impurity position $\bm{r}_a$, and the second term describes the Zeeman coupling with the magnetic field ${\bf B} ({\bf r}_a) $.
Thus, the \emph{ac} electric field couples to the  spins indirectly, by inducing an \emph{ac} supercurrent, which in turn, generates both the magnetic field and the spin-density of the itinerant electrons. 

Below, we develop a microscopic theory of the coupling mechanism between the  spin density of the itinerant electrons in disordered superconductors induced by the supercurrent, and the impurity spins, (the first term in Eq~\eqref{eq:Heff}). We evaluate the dissipative part of the \emph{ac} conductivity associated with electromagnetic field absorption by the system of localized spins. Using the fluctuation-dissipation theorem, we obtain the equilibrium correlation function of the fluctuations of supercurrent density.

The reason why the coupling of the \emph{ac} electric field ${\bf E}(t) = \mathrm{Re}\left(
{\bf E}_{\omega} e^{-i \omega t}\right)$ to the impurity spins  in superconductors is much stronger than that in insulators is that to linear order in ${\bf E}(t)$, the effective magnetic field acting on the magnetic impurities is proportional to the
superfluid momentum ${\bf p}_s (t)$. The latter is given by ${\bf p}_s (t) = \frac{\hbar}{2} \left( \bm{\nabla} \chi - \frac{2e}{\hbar c} \bf{A}(t) \right)$, where $\chi$ is the phase of the order parameter and $\bf{A}(t)$ is the vector potential.
Therefore, ${\bf H}_{\mathrm{eff},a}(t)$ in Eq.~\eqref{eq:Eq_1} may be expressed in the form
\begin{equation}
{\bf H}_{\mathrm{eff},a}(t) = \hat{\zeta}_a \,  {\bf p}_{s}(t) ,
\end{equation}
where $\hat{\zeta}_a$ is a rank-two sensitivity tensor, which has a dimension of velocity and characterizes the linear-in-${\bf p}_s (t)$ breaking of the Kramers degeneracy of the $a$-th impurity. 

The superfluid momentum ${\bf p}_{s}(t)$ and the
electric field are related via the condensate acceleration equation
\begin{equation}
\frac{d{\bf p}_{s}}{dt} = e{\bf E}(t) . 
\end{equation}
Thus, the electric field 
creates the effective \emph{ac} magnetic field, ${\bf H}_{\mathrm{eff},a} (t) = \mathrm{Re} \left( {\bf H}_{\mathrm{eff},a} e^{- i \omega t} \right)$ of the form
\begin{equation}\label{eq:Heffi}
{\bf H}_{\mathrm{eff},a} = i \hat{\zeta}_a\frac{e {\bf E}_{\omega}}{ \omega}  ,
\end{equation}
which is inversely proportional to frequency $\omega$. 

In disordered superconductors, the sensitivity tensors $\hat{\zeta}_a$ are random quantities because both the spin density of the itinerant electrons and the impurity $g$-factors in Eq.~\eqref{eq:Heff} exhibit strong mesoscopic fluctuations. To separate these contributions, we represent the sensitivity tensor as a sum of two terms, $\hat{\zeta}_a = \hat{\zeta}_a^{(1)} + \hat{\zeta}_a^{(2)} $, which correspond to the first and second term in Eq.~\eqref{eq:Heff}, respectively.  
At $\hbar\omega\ll \Delta$,  ${\bf H}_\mathrm{eff}$ turns out to be much larger
than the direct Zeeman coupling of the \emph{ac} magnetic field to the impurity spins.

In the following we assume that the superconductors are in the diffusive regime $l\ll \xi$. Here $\xi=\sqrt{\hbar D/ \Delta}$ is the zero-temperature superconducting coherence length, $l$ is the electronic mean free path associated with non-magnetic impurities, $D=lv_{F}/3$ is the diffusion coefficient, and $v_{F}$ is the Fermi velocity.
To be concrete, we consider the case where the thickness of superconducting samples $d$ is greater than the coherence length and smaller than the London penetration length of the magnetic field, $\xi< d< \lambda_{L}$. In this case, the superfluid momentum may be considered spatially uniform.
We also assume that the frequency $\omega$ of the external field is small, $\hbar \omega \ll \Delta$. 

Let us first discuss the properties of $\hat{\zeta}_a^{(1)}$, which is associated with the spin density of the itinerant electrons,  first term in Eq.~\eqref{eq:Heff}. Using the second-quantized form of the spin density operator, 
$\hat{\bm{ s}}({\bf r}, t) = \psi^\dagger_\alpha ({\bf r},t) \bm{\sigma}_{\alpha \beta}\psi_\beta ({\bf r},t)$, where $\bm{\sigma}_{\alpha \beta}$ is the vector of Pauli matrices and $\psi^\dagger_\alpha ({\bf r})$ and $\psi_\beta ({\bf r})$ are the electron creation and annihilation operators, we express the expectation value of the spin density in terms of the equal-time electron Green's functions, 
\begin{align}
\label{eq:p_s_dependent_exchange_field}
   {\bf s}({\bf r}, t) \equiv  &      \, \langle \hat{\bm{ s}}({\bf r}, t) \rangle =   \bm{\sigma}_{\alpha \beta }\left\langle\psi^\dagger_\alpha ({\bf r}, t) \psi_\beta ({\bf r},t)  \right\rangle.
\end{align}
Here $\langle \ldots \rangle$ denotes the quantum and thermal averaging.
At low frequencies, the time-dependent expectation value of the spin-density operator in Eq.~\eqref{eq:p_s_dependent_exchange_field} is given by the equilibrium spin density for the instantaneous value of the superfluid momentum $\bm{p}_s (t)$. In the absence of supercurrent, $\bm{p}_s =0$, this expectation value vanishes. 
Generation of spin density  (pseudovector) by supercurrent (polar vector), $\langle \hat{\bm{s}} ({\bf r}) \rangle  \sim {\bf p}_{s}$, 
requires not only spin-orbit coupling but also breaking of the inversion symmetry.
We restrict our consideration to disordered centrosymmetric superconductors, where the necessary inversion symmetry breaking is caused by a random distribution of nonmagnetic impurities around the impurity spins. As we show, in this case the effect becomes especially strong. The reason is that in disordered conductors, the spatial form of the wavefunctions of the itinerant electrons is determined by the quantum interference of electron waves scattered from multiple impurities. 
This interference pattern is very sensitive to uniform supercurrent~\cite{spivak1988mesoscopic} which, in the presence of spin-orbit coupling, induces random disorder-specific  fluctuations of spin density of the  itinerant electrons, $\bf s({\bf r})$. 
The spin density fluctuations are characterized by zero average $\overline{\bf s(r)} = 0$ (we denote averaging over disorder by overline) and nonzero correlation function
\cite{spivak1991mesoscopic}
\begin{eqnarray}\label{eq:spin_correlator}
\overline{ s_i({\bf r}) s_j({\bf r}')} = \delta_{ij} \, \overline{ s^{2}}
\times 
\left\{ 
\begin{array}{cc}
1 ,  &   |{\bf r-r'}|\sim \lambda_{F},  \\
\frac{\lambda^2_{F}}{|{\bf r-r'}|^2},  &    l>|{\bf r-r'}|>\lambda_{F},
\end{array}
\right. 
\nonumber \\
\overline{  s^{2}} \sim \frac{\xi}{l} \left(\frac{\hbar  j_{s} }{eD p_{F}}\right)^{2} \times
\left\{ 
\begin{array}{cc}
 1 ,   &   l_{\text{so}}<\xi,  \\
\hbar/ (\Delta \tau_{\text{so}}),  &    l_{\text{so}}>\xi.
\end{array}
\right.
\end{eqnarray}
Here $\delta_{ij}$ is the unit tensor, and $l_{\text{so}} =\sqrt{D\tau_{\text{so}}}$ and $\tau_{\text{so}}$
are the spin-orbit relaxation length and time, respectively, for the itinerant electrons due to scattering by nonmagnetic impurities. Finally, 
\begin{align}
\label{eq:supercurrent_p_s}
    \bm{j}_s (t) = & \, - \frac{e N_s}{m} {\bf p}_s (t)  = \frac{e^2 N_s}{m \omega } \mathrm{Im} \left(  {\bf E}_\omega e^{- i \omega t} \right) 
\end{align}
is the super-current density, and $N_s$ is the superfluid density. 

\begin{figure}
    \centering
    \includegraphics[width=0.7\linewidth]{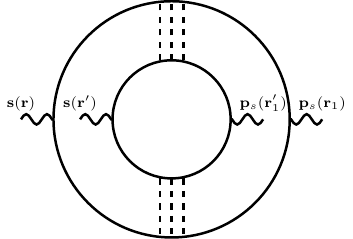}
    \caption{The Feynman diagram for the spin correlation function induced by the supercurrent.
The dashed lines represent disorder averaging, and
solid lines represent the matrix electron Green’s functions in the Gor'kov-Nambu space.}
    \label{fig:placeholder}
\end{figure}

One obtains Eq.~\eqref{eq:spin_correlator} using the standard impurity diagram technique (see for example Ref.~\cite{abrikosov1959application}). The Feynman diagram describing the disorder-averaged correlation function of the spin densities is shown in Fig.~\ref{fig:placeholder}.  Note that ${\bf s}({\bf r})$ has a random direction. Furthermore, for short spin-orbit relaxation length, $l_{\text{so}}<\xi$, the amplitude of spin fluctuations becomes independent of $\tau_{\text{so}}$.
Therefore, the different elements of the random tensor $\hat{\zeta}_a^{(1)}$, have a variance of the same order, $\overline{\zeta_1^2}$, which is given by
 \begin{equation}
    \overline{\zeta^{2}_{1}} \sim \mathcal{J}^2 \frac{\xi}{l}\left(\frac{\hbar}{mDp_F}N_s\right)^2\times\begin{cases}
  1   &,  \,\,\ l_{so}<\xi,  \\
  \hbar/ \left(\Delta \tau_{\text{so}}\right)  &,   \,\,\ l_{so}>\xi.
\end{cases}
\end{equation}

Let us now estimate the second contribution to the sensitivity tensor, $\hat{\zeta}_a^{(2)}$, which is associated with the magnetic field induced in the superconductor by the \emph{ac} electric field. We assume the film to be perpendicular to the $z$-axis. 
The magnetic field in the film depends on the $z$-coordinate. This dependence is obtained using the Maxwell's equation $\partial_z {\bf B} (t,z) = -\hat{z}\times \left[ \frac{4\pi}{c}  \bm{j}_s (t) + \frac{1}{c} \partial_t \bf{E}(t)\right]$, where $\hat{z}$ is the unit vector normal to the film. 
As follows from Eq.~\eqref{eq:supercurrent_p_s},  for thin films, $d\ll \lambda_L$, where the electric field and the current density are independent of $z$, the magnetic field is given by ${\bf B}(z,t)= \frac{z \omega }{c}    \mathrm{Re} \left( i \left[ \frac{\Tilde{\omega}_p^2}{\omega^2}-1\right] \hat{z} \times \bf{E}_\omega  e^{- i \omega t}\right)$. Here we introduced an effective plasma frequency $\Tilde{\omega}_p^2 = \frac{4\pi N_s e^2}
{m}$. 
At low frequencies, the  magnetic field becomes inversely proportional to $\omega$, and the characteristic Zeeman energy in Eq.~\eqref{eq:spin_hamiltonian} may be estimated as,  
\begin{align}\label{eq:B_Zeeman_estimate}
g_a\mu_{\mathrm{B}} B \sim & \, g_a E_\omega d \frac{|e| \hbar}{2 m c^2} \frac{  \tilde{\omega}_p^2  }{\omega}.
\end{align}
Thus, the typical value of  the second contribution of the sensitivity tensor may be estimated as 
\[
\zeta_{2} \sim g_a d\hbar \tilde{\omega}^{2}_{p}/2mc^{2}.
\]The impurity $g$-factor has  spatial fluctuations $\delta g_{a}=g_{a}-g_{0}$ (with $g_{0}$ being the $g$-factor in a pure sample), which are of mesoscopic origin: an impurity spin embedded into a disordered metal induces random Friedel oscillations of magnetization, which control the value of $\delta g_{a}$. We use the procedure similar to that of Ref.~\cite{zyuzin1986friedel} to estimate the relative amplitude of $g$-factor fluctuations, 
$\sqrt{ \overline{\left(\delta g_{a} \right)^{2}}}/g_{0} \sim 1/(k_{F}l)^{3/2}$.

In linear response, the individual impurity spins immersed into a TRS-invariant system can absorb the external electromagnetic field only to the extent that the energies of spin states are broadened either by spin-phonon processes or by the hyperfine interaction with the nuclear spins. Indeed, in the absence of coupling between the impurity spins, their energy levels are doubly degenerate by Kramers symmetry. Therefore, the absorption rate is finite only in a very narrow frequency interval of the order of the broadening 
(see for example Ref.~\cite{glazov2012spin}).
At higher frequencies, the absorption is associated with the RKKY interaction between the impurity spins, which lifts the degeneracy of the energy levels of the spin subsystem. The RKKY interaction is generated in second order in the exchange interaction with the spin density of the itinerant electrons.  At distances smaller than the coherence length,
$|{\bf r}_{a}-{\bf r}_{b}| < \xi$, 
the RKKY exchange coupling is  essentially the same as that in a normal metal \cite{ruderman1954indirect,kasuya1956electrical,yosida1957magnetic}. At $|{\bf r}_{a}-{\bf r}_{b}|>\lambda_{F}$, it has the form~\footnote{ In the presence of spin-orbit coupling, the RKKY interaction of the impurity spins in Eq.~\eqref{eq:RK} in general becomes anisotropic, $ J^{ij}_{ab} S^i_a S^j_b$, with $i$ and $j$ being the Cartesian indices. However, the anisotropy of the RKKY exchange does not affect our conclusions.} 
\begin{subequations}\label{eq:RKKY}
\label{eq:RK}
\begin{align}
\label{eq:RK_a}
\hat{H}_{\mathrm{RKKY}} = & \, \frac{1}{2}\sum_{ab} J_{ab} \, \hat{\bm{S}}_a\cdot \hat{\bm{S}}_b,  \\
    \label{eq:RK_b}
J_{ab}=&\,  \frac{\mathcal{J}^2\nu}{4\pi}\frac{\cos(2k_F|{\bf r}_{a}-{\bf r}_{b}|+\phi({\bf r}_{a}-{\bf r}_{b}))}{|{\bf r}_{a}-{\bf r}_{b}|^{3}},
\end{align} 
\end{subequations}
where $\nu$ is the electron density of states in the normal metal.
Here the phase  $\phi({\bf r}_{a}-{\bf r}_{b})$ 
is a random sample-specific quantity \cite{zyuzin1986friedel}. 
At larger distances, $|{\bf r}_{a}-{\bf r}_{b}| >\xi$, the RKKY coupling becomes suppressed by an additional factor 
$\exp(-2|{\bf r}_{a}-{\bf r}_{b}| /\xi)$. 
The full Hamiltonian of the spin subsystem is 
\begin{equation}
\hat{H}= \hat{H}_{\mathrm{RKKY}}+\hat{H}_\mathrm{S}.
\end{equation}
The criterion that the paramagnetic impurities do not affect the superconducting gap is $\Delta\tau_{s}/\hbar\gg 1$, where $1/\tau_{s} \propto n_s$ is the spin-flip rate for the itinerant electrons, which is associated with scattering on paramagnetic impurities.
This leaves us with an interval of the localized spin concentration of several orders of magnitude,
\begin{equation}\label{eq:n_{s}inequality}
\frac{1}{\xi^{3}
}< 
n_{s} \ll 
\frac{\Delta}{\hbar\lambda_{F}^{2}v_{F}}.
\end{equation}
Here we assumed that the scattering length for the spin-flip scattering of electrons on the paramagnetic impurities is of the order of the Fermi wavelength $\lambda_{F}$.

In the absence of an electromagnetic field, the typical energy of exchange interaction between two spins separated by a  typical distance $ n_{s}^{-1/3}$ is of the order of 
\begin{equation}
J\sim \frac{\mathcal{J}^{2}\nu}{4\pi}n_{s}.
\end{equation}
At temperatures lower than the spin glass transition temperature,
$T<T_{SG}\sim J$, the system of spins described by the Hamiltonian $\hat{H}_\mathrm{RKKY}$ is in the spin glass state. 
In this article, we focus on the regime of relatively high temperatures,
\begin{equation}
T_c \gg T>T_{SG}.     
 \end{equation}
We assume that in this temperature interval the spin system is ergodic, and its energy propagates in space via spin diffusion.  In this case, the broadening of the energy levels of individual spins is of order $J$. 

Let us now evaluate the dissipative part of the \emph{ac} conductivity caused by the coupling of supercurrent to the impurity spins. We focus our consideration on films of sufficiently small thickness, where the ratio
\begin{equation}\label{eq:ratio}
\frac{\zeta_{1}}{\zeta_{2}}\sim \frac{mc^{2}}{e^{2}k_{F}}
\frac{\lambda^{2}_{F}}{ld} \left(\frac{\xi}{l}\right)^{1/2}
\end{equation}
is larger than unity~\footnote{To get this estimate we assumed that $l_{so}<\xi$ and used $\mathcal{J}\sim \epsilon_{F}\lambda_{F}^{3}$, with $\epsilon_{F}$ being the Fermi energy. }. Thus,
the coupling of the impurity spins to the spin density of the itinerant electrons, (the first term in 
Eq.~\eqref{eq:Heff}), gives the dominant contribution to the low frequency dissipative conductivity
\footnote{
For superconductors belonging to the cubic crystalline system the RKKY exchange is isotropic, and $\hat{H}_{\text{RKKY}}$ in Eq.~\eqref{eq:RKKY} commutes with the total spin $\sum_a {\bf S}_a$. In this case, in the absence of $g$-factor variations, $g_{a}=g_{0}$, a spatially uniform \emph{ac} magnetic field does not produce energy dissipation.   
As a result, the ratio of contributions to the dissipative conductivity due to the first  and the second terms in Eq.~\eqref{eq:Heff} is enhanced compared to Eq.~\eqref{eq:ratio} by a large parameter $C \sim  \max\left[g/\sqrt{\overline{ (\delta g_{a})^{2}}}; \frac{d}{\tilde{r}}\right]\gg 1$, reflecting the variation of the $g$-factors  and  spatial nonuniformity of the magnetic field.
Here $\tilde{r} $ is the typical distance between the impurity spins relevant to the problem. }.

Generally, there are two theoretical paradigms of the absorption of the electromagnetic field: 
resonance absorption and the relaxation (Debye) mechanism (see for example Refs. \cite{JACKLE1976365,HUNKLINGER1986265}), giving two contributions to the dissipative part of the conductivity $\sigma(\omega)=\sigma_{R}(\omega)+\sigma_{D}(\omega)$. 
The two contributions can be uniquely distinguished in the case where the splitting between the  relevant energy states is larger than their broadening, which is of order $J$. 

Let us first estimate the part of the conductivity $\sigma_{R}(\omega)$ associated with the resonance absorption.
At frequencies $\hbar \omega>J$ the resonance  absorption is associated with the transitions  between the ground and excited states of the rare pairs of spins separated by a short distance $r^{*}(\omega)\ll n_{s}^{-1/3}$.   The strong RKKY exchange creates the energy splitting $E_{\mathrm{spl}}=\hbar \omega $ between the singlet and the triplet manifold for this pair~\footnote{For anisotropic exchange, the triplet will be split into a singlet and a doublet, but this does not affect our conclusion.},
\begin{equation}
\label{eq:E_spl_r_star}
E_{\mathrm{spl}}\left(r^{*}\right)\sim \frac{J}{(r^{*})^{3}n_{s}}\gg J,
\end{equation}
which is much larger than their broadening. Thus, the energies of the spin state of the close pairs are relatively well defined in this regime. 
For uncorrelated impurities, the probability density to find a pair of spins of the size $r^{*}\ll n_s^{-1/3}$  is proportional to $ n_{s}\cdot (r^{*})^{2}$.   Using Eq.~\eqref{eq:E_spl_r_star}
we get the probability density for the energy splitting $E_{\mathrm{spl}}$ between the singlet and triplet manifolds for close pairs,
\begin{equation}
\label{eq:p_E}
p (E_{\mathrm{spl}}) \sim \frac{J}{E_{\mathrm{spl}}^2 }, \,\,\,\ E_{\mathrm{spl}} \geq J .
\end{equation}
The resulting resonance absorption contributing to the conductivity is
\begin{align}
\label{eq:sigmaR}
\sigma_{R} (\omega) \sim & \, \frac{e^{2}n_{s}\overline{ \zeta^{2}}}{{\hbar}\omega^2}\int d E_{\mathrm{spl}} \,  
p(E_{\mathrm{spl}})   f\left(E_{\mathrm{spl}}\right)E_{\mathrm{spl}}   \, \delta(\hbar\omega - E_{\mathrm{spl}}) 
\nonumber 
   \\
   \sim &\, \frac{e^{2}n_{s} \overline{\zeta^{2}}}{{\hbar}}  \begin{cases}
    \frac{J}{T\omega^2} , &T> \hbar\omega \geq J ,\\
    \frac{J}{{\hbar}\omega^3}, &{\hbar}\omega > T >J,
    \end{cases}
\end{align}  
where $f\left(E_{\mathrm{spl}}\right)$ is the difference in the equilibrium occupation numbers of the singlet and the triplet manifold of the spin pair with the energy splitting $E_{\mathrm{spl}}$.

At $\hbar\omega\lesssim J$  the contribution of resonance absorption to conductivity   saturates to a frequency-independent value,  
\begin{equation}\label{eq:sigma_{R}0}
\sigma_{R} (\omega) \sim e^2n_s \overline{\zeta^2} \frac{{\hbar}}{TJ} , \,\quad \hbar\omega\lesssim J.
\end{equation}
In this regime $\sigma_R$ is dominated by the typical spin pairs.

We now turn to the discussion of the Debye contribution to the \emph{ac} conductivity $\sigma_{D}(\omega)$. Its mechanism can be understood as follows~\cite{debye1929polar}. In the adiabatic approximation,
$\hbar\omega<E_{\mathrm{spl}}$, the effective magnetic field modulates the energies of the spin system, creating a non-equilibrium population of the energy levels. The relaxation of the nonequilibrium level population is accompanied by the entropy production and produces a contribution to the dissipative conductivity $\sigma_{D}$.

Usually, the Debye contribution to the \emph{ac} conductivity $\sigma_{D}(\omega)$ becomes dominant at low frequencies in systems that are characterized by a hierarchy of relaxation times. In this case, the conductivity is proportional to the longest relaxation time.
We will argue that such a hierarchy of relaxation times arises when a static in-plane magnetic field ${\bf B}_{dc}$ is applied to the system. In this case, the Debye contribution to the conductivity can be much larger than the resonant one, $\sigma_{D} \gg \sigma_{R}$, which implies a large positive magnetoconductance.   
In this situation, $\sigma_D$ is also dominated by closely spaced pairs of spins. The reason is that because of the large energy splitting between the triplet and singlet manifolds for such pairs, the full relaxation of the nonequilibrium distribution of energy levels involves long relaxation times.  

Let us consider the adiabatic evolution of the energy levels of the closely spaced spin pairs in the presence of a \emph{dc} magnetic field ${\bf B}_{dc}$. The conductivity $\sigma_D (\omega)$ arises from the motion of levels linear in the \emph{ac} field ${\bf H}_{\mathrm{eff},a}$. The linear in  ${\bf H}_{\mathrm{eff},a}$ breaking of the degeneracy of the triplet manifold arises even at ${\bf B}_{dc}=0$. However, in this case the average energy of the triplet states does not shift, and the total
equilibrium population of the triplet manifold remains unchanged to linear order in  ${\bf H}_{\mathrm{eff},a}$.
In the presence of ${\bf B}_{dc}$, the average energy of the triplet manifold shifts proportionally to ${\bf H}_{\mathrm{eff},a}\cdot{\bf B}_{dc}$. 
Thus, at ${\bf B}_{dc} \neq 0$ the spectrum evolution creates a population imbalance between the singlet and triplet manifolds. In the absence of relaxation between the triplet and singlet manifold, this imbalance may be estimated as
\begin{equation}
\label{eq:delta_f_adibatic}
\delta f_{0}\sim \frac{\mu_B{\bf B}_{dc} \cdot  {\bf H}_{\mathrm{eff},a}}{T^{2}}.
\end{equation}    
Here we assumed that $E_{\mathrm{spl}}\lesssim T$. 

\begin{figure}
    \centering
    \includegraphics[width=0.9\linewidth,trim={0 3cm 0 0},clip]{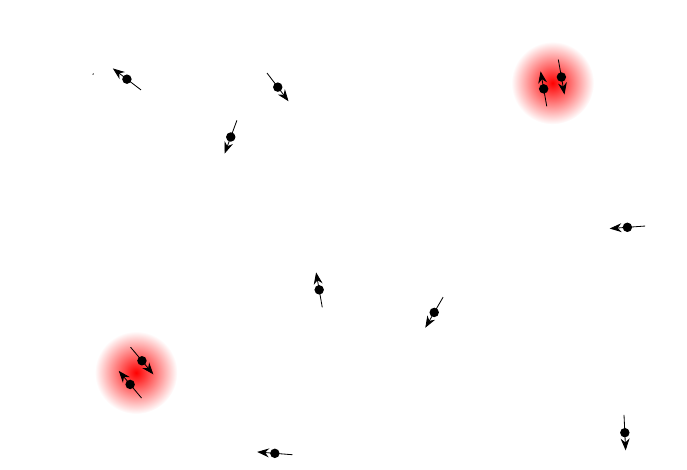}
    \caption{Schematic depiction of  paramagnetic impurities in a superconductor. Two rare ``resonance" pairs, which have close singlet-triplet energy splittings but are separated by a large distance, are shaded in red.}

    \label{fig:rarepairs}
\end{figure}

Equilibration within the triplet manifold involves small energy transfer and is characterized by a short relaxation time on the order $\hbar/J$. 
In contrast, equilibration between the singlet and triplet manifolds of the closely spaced spin pairs involves processes with energy transfer of order $E_{\mathrm{spl}} \gg J$. This cannot be achieved by spin-flip processes involving spins located distances of order $n_{s}^{-1/3}$ from the close pair because such processes do not conserve energy. 
Thus, equilibration between the triplet and singlet manifolds via the spin-flip processes involves rare ``resonant" pairs with the same (within the broadening energy interval $J$) energy splitting $E_{\mathrm{spl}}$. Two such resonant pairs are shown in red in Fig.~\ref{fig:rarepairs}.

The density of such rare resonant pairs is of order $n_{p} = n_s J p(E_{\mathrm{spl}})$. The  matrix element for such resonant spin-flips is $
J_{p}\sim J \frac{n_{p}}{n_{s}}
%\sim J (n_s (r^*)^3)^2 
 \sim J\frac{J^2}{E^2_{\mathrm{spl}}}$
and the corresponding relaxation time is of order  
 \begin{equation}
 \label{eq:tau_E}
\tau(E_{\mathrm{spl}})\sim \frac{\hbar J}{J_{p}^{2}} 
 \sim \frac{\hbar}{J} \frac{E_{\mathrm{spl}}^4}{J^4}\gg \frac{\hbar}{J}.
 \end{equation}
A similar situation was considered in the context of dielectric glasses in Ref.~\cite{burin1995low}.

To evaluate the Debye contribution to conductivity, $\sigma_{D}$,  we follow the standard steps (see e.g. Refs.~\onlinecite{JACKLE1976365,HUNKLINGER1986265}). 
In the presence of relaxation, the amplitude of the population imbalance between the singlet and triplet manifolds changes from Eq.~\eqref{eq:delta_f_adibatic} to $\delta f (\omega) \sim  \delta f_0 \frac{  \omega  \tau (E_{\mathrm{spl}}) }{ i +  
\omega \tau (E_{\mathrm{spl}}) } $. Since the entropy production rate is proportional to $|\delta f (\omega)|^2/\tau(E_{{\mathrm{spl}}})$,   
we  get 
\begin{align}
\label{eq:Pdebye}
 \sigma_{D} (\omega, {\bf B}_{dc}) & \sim   e^{2} n_{s}\overline{\zeta^{2}} \mu_{\mathrm{B}}^{2} B_{dc}^{2} \nonumber \\
& \times \int_{J}^{E_\mathrm{max}}   \frac{p(E_{\mathrm{spl}})  \tau ( E_{\mathrm{spl}})\,  d  E_{\mathrm{spl}}}{T^{3} \left[1+\omega^2 \tau^2( E_{\mathrm{spl}}) \right] }.
\end{align}
Here, the upper limit of the integration, $E_\mathrm{max}=\min [\sqrt{n_s \xi^3} J,T]$, 
is determined by the requirements that  
$n_{p}^{-1/3}< \xi$, and $E_\mathrm{spl}<T$. 
The main difference between Eq.~\eqref{eq:Pdebye} and the standard expression for the Debye conductivity in dielectrics, Ref.~\cite{JACKLE1976365,HUNKLINGER1986265}, is that Eq.~\eqref{eq:Pdebye} does not vanish at $\omega=0$.

Substituting Eqs.~\eqref{eq:p_E} and \eqref{eq:tau_E}  into Eq.~\eqref{eq:Pdebye} we obtain 
\begin{equation}
\label{eq:sigma_results1}
    \sigma_D(\omega,{\bf{B}}_{dc}) \sim \frac{\hbar e^2 \overline {\zeta^2} \mu^2_B B^2_{dc}}{JT^3}\begin{cases}
        \left(\frac{J \tau_\mathrm{max}}{\hbar} \right)^{3/4}, \frac{\hbar}{\tau_\mathrm{max} }> \hbar \omega\\
        \left(\frac{J}{\hbar \omega}\right)^{3/4}, \frac{\hbar}{\tau_\mathrm{max}}< \hbar \omega < J,
    \end{cases}
\end{equation}
where 
\begin{equation}
\tau_\mathrm{max} = \min\left[\frac{{\hbar}T^4}{J^5},\frac{{\hbar}n_s^{2}\xi^6}{J} \right]
\end{equation}
is the maximum relaxation time in the system.

Similar to the resonance contribution to the conductivity, Eq.~\eqref{eq:sigma_{R}0}, the frequency dependence of the Debye contribution, Eq.~\eqref{eq:sigma_results1}, saturates at small frequencies. 
In this regime, it exceeds $\sigma_R(0)$ when 
\begin{equation}\label{eq:Hdc}
 \mu_{\mathrm{B}}B_{dc} > J\sqrt{\frac{J}{T}}.
\end{equation}
Therefore, if the condition \eqref{eq:Hdc} holds, the magnetoconductivity of the system described by Eqs.~\eqref{eq:Pdebye}, \eqref{eq:sigma_results1}, becomes larger than the conductivity at zero magnetic field, Eq.~\eqref{eq:sigma_{R}0}. 
The quadratic dependence of the magnetoconductivity on $B_{dc}$ in Eq.~\eqref{eq:Pdebye} was obtained using the second order of perturbation theory. This is valid if the characteristic energy splitting of triplet states, $\mu_{B}B_{dc}$,  is smaller than the characteristic energy splitting  $E_{\mathrm{spl}}$ that gives the main contribution to the integral in Eq.~\eqref{eq:Pdebye}. For $\tau_\mathrm{max} = \hbar T^4/J^5$, this condition corresponds to  $\mu_{B}B_{dc}<T$.

At ${\bf B}_{dc}=0$ and in the linear order in the \emph{ac} field,  the population imbalance between the triplet and  singlet manifolds of closely spaced pairs is not generated. 
Therefore, the relaxation mechanism is characterized by a single short relaxation time of order $\hbar /J$. The resulting contribution to the low frequency conductivity is of the same order as the resonance contribution $\sigma_R$ in Eqs.~\eqref{eq:sigmaR}, and \eqref{eq:sigma_{R}0}. 

It follows from Eq.~\eqref{eq:sigma_results1} that at $B_{dc} =0$ the threshold for nonlinearity of the dissipative conductivity should be anomalously small. 
Indeed, for a fixed \emph{ac} electric field $E_\omega$, the amplitude of supercurrent oscillations is proportional to $1/\omega$. Therefore, at low frequencies  the response of the system becomes nonlinear at very
small $E_\omega$. There are two requirements for the validity of the linear response results, Eqs.~\eqref{eq:sigmaR}, \eqref{eq:sigma_{R}0}, and \eqref{eq:sigma_results1}; (i)  the amplitude of the oscillations of the superfluid momentum must be smaller than the critical one $p_{sc}\sim \hbar/\xi$, which yields $eE/\omega\ll \hbar/\xi$;  (ii)   $H_\mathrm{eff}\ll \mu_{\mathrm{B}} B_{dc}$. 
In the nonlinear regime, the dissipative conductivity at $B_{dc}=0$ may be estimated by substituting $\mu_{\mathrm{B}}{\bf B}_{dc} \rightarrow {\bf H}_\mathrm{eff}$ in Eq.~\eqref{eq:sigma_results1}.

At $T>T_{SG}$, all conventional theorems of statistical mechanics  are valid in superconductors with paramagnetic impurities. One can use the fluctuation-dissipation theorem (FDT) \cite{lifshitz2013statistical} to relate the spectral density of fluctuations of the total current through the system to the dissipative conductivity
\begin{equation}\label{eq:S}
%S_{s}(\omega)=
\int dt \overline{ \left\langle \delta I_{s}(t) \delta I_{s}(0) \right\rangle} e^{i\omega t} = \hbar \omega \sigma(\omega) \coth \frac{\hbar \omega}{2T} \, d .
\end{equation}
Thus,  the spectral power of current fluctuations also saturates to a nonzero value at low frequencies, $\omega\rightarrow 0$.
The subscript $s$ in Eq.~\eqref{eq:S} indicates that the system is in the superconducting state, and $\delta I_{s}(t)$ denotes fluctuations of the total current through a two-dimensional square sample with thickness $d$. These fluctuations can be detected by the measurements of magnetic flux noise or fluctuations of the Faraday rotation of light polarization (see, for example, Ref.~\cite{glazov2018electron}).

The authors are grateful to M. Feigelman, M.M. Glazov and E.L. Ivchenko  for helpful discussions. The work of B.S. and S-B. C. was supported by the
DARPA grant GR049687, A.A. was supported by the
NSF grant DMR-2424364, AB is supported by NSF grant CHE-2201027, and NSF-BoR EPSCORE LINK Program, award 20130015223. 

\bibliographystyle{unsrt}
\bibliography{ref} 

\end{document}